\documentclass[reprint, lettepaper, superscriptaddress]{revtex4-1}
\usepackage{graphicx}
\usepackage{amsmath}
\usepackage{esint}
\usepackage{verbatim}
\usepackage{color}
\usepackage{SIunits}
\usepackage{hyperref}

\begin{document}

\title{Nanowire photonic crystal waveguides for single-atom trapping and strong light-matter interactions}

\author{S.-P. Yu}
\thanks{These authors contributed equally to this work}
\author{J. D. Hood}
\thanks{These authors contributed equally to this work}
\affiliation{Norman Bridge Laboratory of Physics 12-33}
\affiliation{Institute for Quantum Information and Matter}
\author{J. A. Muniz}
\author{M. J. Martin}
\affiliation{Norman Bridge Laboratory of Physics 12-33}
\affiliation{Institute for Quantum Information and Matter}
\author{Richard Norte}
\affiliation{Institute for Quantum Information and Matter}
\affiliation{Thomas J. Watson, Sr., Laboratory of Applied Physics 128-95, California Institute of Technology, Pasadena, CA 91125, USA}
\author{C.-L. Hung}
\affiliation{Norman Bridge Laboratory of Physics 12-33}
\affiliation{Institute for Quantum Information and Matter}
\author{Se\'{a}n M. Meenehan}
\author{Justin D. Cohen}
\affiliation{Institute for Quantum Information and Matter}
\affiliation{Thomas J. Watson, Sr., Laboratory of Applied Physics 128-95, California Institute of Technology, Pasadena, CA 91125, USA}
\author{Oskar Painter}
\email{opainter@caltech.edu}
\affiliation{Institute for Quantum Information and Matter}
\affiliation{Thomas J. Watson, Sr., Laboratory of Applied Physics 128-95, California Institute of Technology, Pasadena, CA 91125, USA}
\author{H. J. Kimble}
\email{hjkimble@caltech.edu}
\affiliation{Norman Bridge Laboratory of Physics 12-33}
\affiliation{Institute for Quantum Information and Matter}

\date{\today}
\begin{abstract}
We present a comprehensive study of dispersion-engineered nanowire photonic crystal waveguides suitable for experiments in quantum optics and atomic physics with optically trapped atoms.  Detailed design methodology and specifications are provided, as are the processing steps used to create silicon nitride waveguides of low optical loss in the near-IR.  Measurements of the waveguide
optical properties and power-handling capability are also presented.   
\end{abstract}
\maketitle

A new frontier for optical physics would become accessible with the integration of atomic systems and nanophotonics, which have made remarkable advances in the last decade \cite{Barclay_thesis, Eichenfield:2009gj, Taguchi:2011ur, Tien2011,Luke:2013uo}. 
Significant progress toward integration of atomic systems with photonic devices has progressed on several fronts, including cavity QED, where atom-photon interactions can be enhanced in micro- and nanoscopic optical cavities \cite{thompson2013b, Vahala:2003cx, Lev:2004bn, Aoki:2006gq, Lepert:2011vv, Volz:2011fl}, and nanoscopic dielectric waveguides, where the effective area of a guided mode can be comparable to atomic radiative cross sections leading to novel photon transport in 1D~\cite{Shen:2005wg,LeKien:2005bo,Dzsotjan:2010bj, Chang:2007fk, Chang:2012co,Asboth:2008jh,Chang:2013bh, Hung2013}, as recently demonstrated in Refs. \cite{Vetsch2010,Dawkins2011,Goban2012, Goban:2013wp}. 

Beyond traditional settings of cavity QED and waveguides, new paradigms emerge by combining atomic physics with photonic crystal waveguides. One- and two-dimensional photonic crystal structures formed from planar dielectrics~\cite{Painter_thesis} offer a configurable platform for engineering strong light-matter coupling for single atoms and photons with circuit-level complexity.  For instance, dispersion-engineered photonic crystal waveguides permit the trapping and probing of ultracold neutral atoms with commensurate spatial periodicity for both trap and probe optical fields that have disparate free-space wavelengths~\cite{Hung2013}.  Such systems can lead to atom-atom interactions efficiently mediated by photons within the waveguide~\cite{John:1990cn, Bhat:2006jm, Shahmoon:2013gk, Douglas:2013tb}. In photonic crystal waveguides, atom-photon coupling can be enhanced near the band-edge via slow-light effects~\cite{Hung2013,Lodahl2013}, and can be tailored to explore quantum many-body physics with atom-atom interactions that can be readily engineered~ \cite{Douglas:2013tb}.

\begin{figure}[btp]
\begin{center}
\includegraphics[width=\columnwidth]{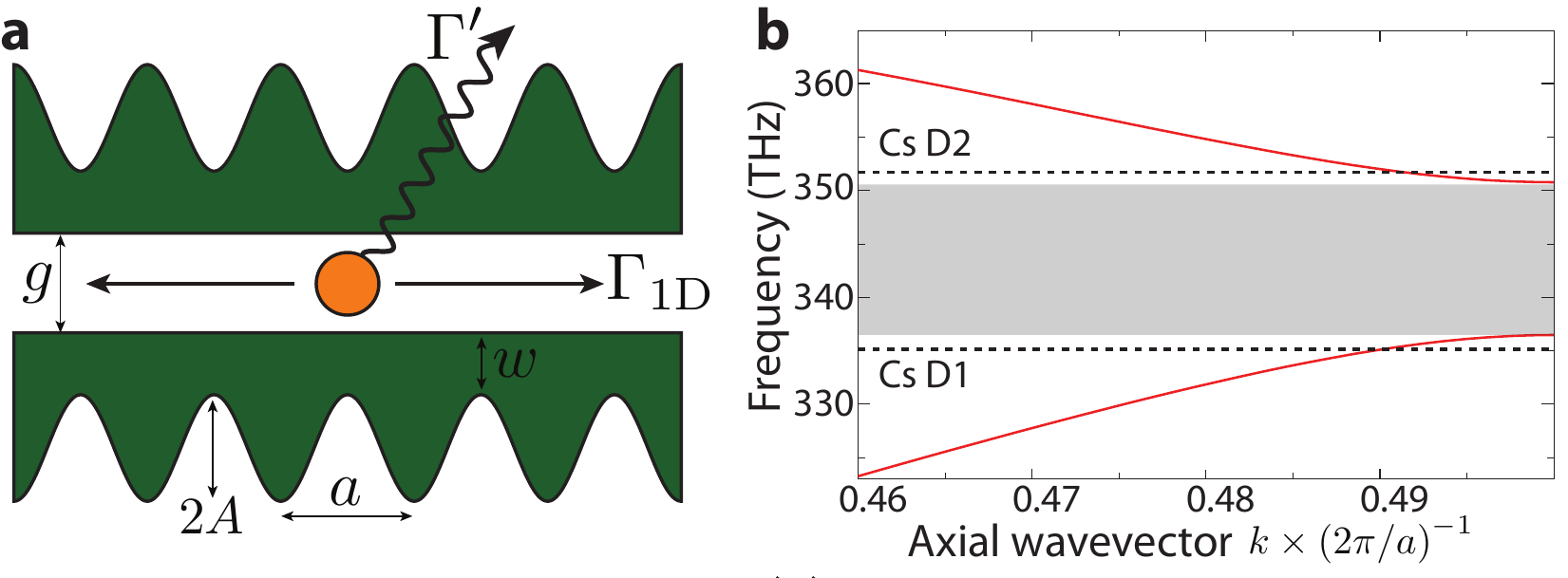}
\vspace{-6mm}
\caption{(a) Schematic of the alligator photonic crystal waveguide (APCW) with dimensional parameters thickness $t=200$~nm, inner waveguide width $w=187$~nm, gap $g=260$~nm, discrete periodicity $a=371$~nm, and sinusoidally-modulated outer waveguide edge with `tooth' amplitude $A=129$~nm. (b) Photonic bandstructure of the fundamental TE-like modes of a nominal alligator waveguide device calculated with the dimensions derived from a typical fabricated device \cite{MPB}. Small adjustments are made to the waveguide parameters within the absolute uncertainty of the SEM ($<5$\%) to obtain better agreement between measured band structures and those computed from SEM images.} \label{fig:wg_design}
\end{center} 
\vspace{-8mm}
\end{figure}

Significant technical challenges exist for developing hybrid atom-photonic systems arising from the following requirements:  (1) The fabrication is sufficiently precise to match waveguide photonic properties to atomic spectral lines; (2) atoms are stably trapped in the presence of substantial Casimir-Polder forces \cite{Hung2013} yet achieve strong atom-field interaction; (3) coupling to and from guided modes of nanophotonic elements is efficient; (4) sufficient optical access exits for external laser cooling and trapping; and (5) optical absorption is low and the net device thermal conductivity is high, permitting optical power handling to support $ \sim \! \! 1$~mK trap depths.  In this Letter, we describe nanowire photonic crystal waveguides that meet these stringent requirements for integration of nanophotonics with ultracold atom experiments. 

The central component of our device is the  `alligator' photonic crystal waveguide (APCW) region shown in Fig.~\ref{fig:wg_design}a. It consists of two parallel Si$_3$N$_4$ waveguides (refractive index $n=2$). This configuration is similar to that proposed in Ref.~\cite{Hung2013} for the trapping of atoms in the gap between the dielectrics, where the atomic spontaneous emission rate into a single guided mode, $\Gamma_{1D}$, can be greatly enhanced with respect to spontaneous emission into all other free-space and guided modes, $\Gamma'$, which is approximately equal to the free-space spontaneous emission rate, $\Gamma_{0}$. Figure~\ref{fig:wg_design}b shows the theoretical optical bandstructure of the TE-like modes (electric field polarized in the plane of the waveguide) for the APCW studied in this work, computed using the MIT Photonic Bands (MPB)~\cite{MPB} software package.  The waveguides are designed such that the Cs D1 ($\nu_1=335.1$~THz) and D2 ($\nu_1=351.7$~THz) transitions are aligned near the lower/`dielectric' ($\nu_D$) and upper/`air' ($\nu_A$) band-edges, respectively.

\begin{figure}[btp]
\begin{center}
\includegraphics[width=\columnwidth]{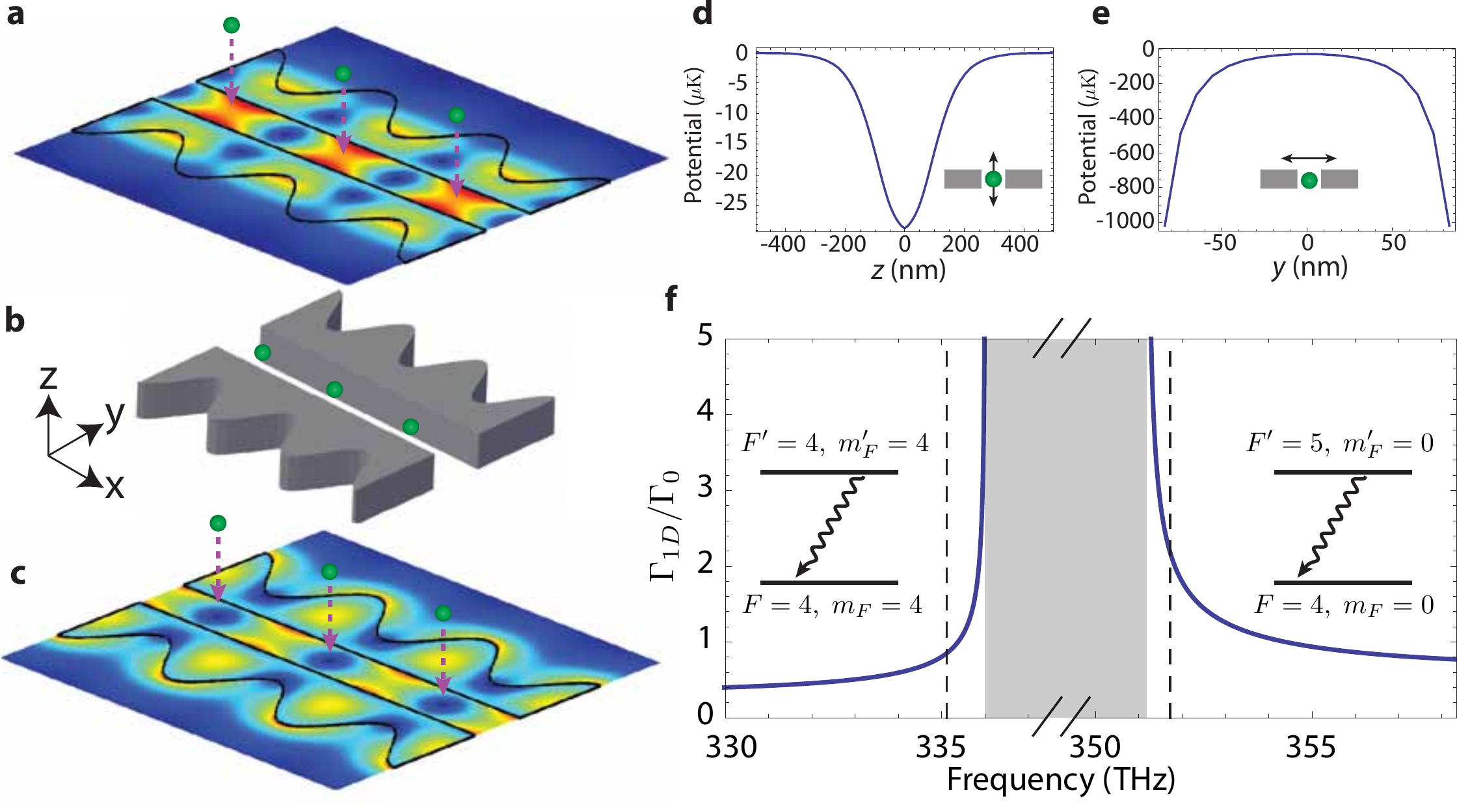}
\caption{Finite-element-method (FEM) simulation of the near-$X$-point guided mode electric field magnitudes $|\vec{E}(\vec{r})|$ in the $x$-$y$ plane for the (a) air band and (c) dielectric band of the even parity TE-like supermodes for the periodic structure shown in (b). The optical frequencies correspond to the Cs D1 and D2 lines, and the corresponding band structure is shown in Fig.~1b.  (d-e) Numerically computed Casimir-Polder potential along directions $(x_m,y_m,z)$ (d) and $(x_m,y,z_m)$ (e) for the dielectric-band trapping mode around minima of the optical trapping potential at $(x_m,y_m,z_m)$ [i.e., the positions of the green spheres in (c)].
(f) Calculated rate of radiative decay $\Gamma_{1\mathrm{D}}$ into the guided mode in (a) for the cases of an initially excited atom trapped at $(x_m,y_m,z_m)$ in an infinite photonic crystal for transitions between atomic levels as depicted in the figure. The shaded area indicates the photonic bandgap region and the dashed lines the Cs D1 and D2 transition frequencies. Here, $\Gamma_0$ is the free-space decay rate.} \label{fig:mode_trap}
\end{center}
\vspace{-8mm}
\end{figure}

As discussed in detail in Ref.~\cite{Hung2013}, the enhanced density of states near the $X$-point band-edge, along with the strong field confinement of the even-parity supermodes in the gap, can be used to create large atom-photon interactions.  Intensity images of the dielectric and air band modes are plotted in Fig.~\ref{fig:mode_trap}a and b, respectively \cite{MPB}. The corresponding enhancement of $\Gamma_{1D}$ is shown in Fig.~2f.

\begin{figure*}[t!]
\begin{center}
\includegraphics[width=1 \textwidth]{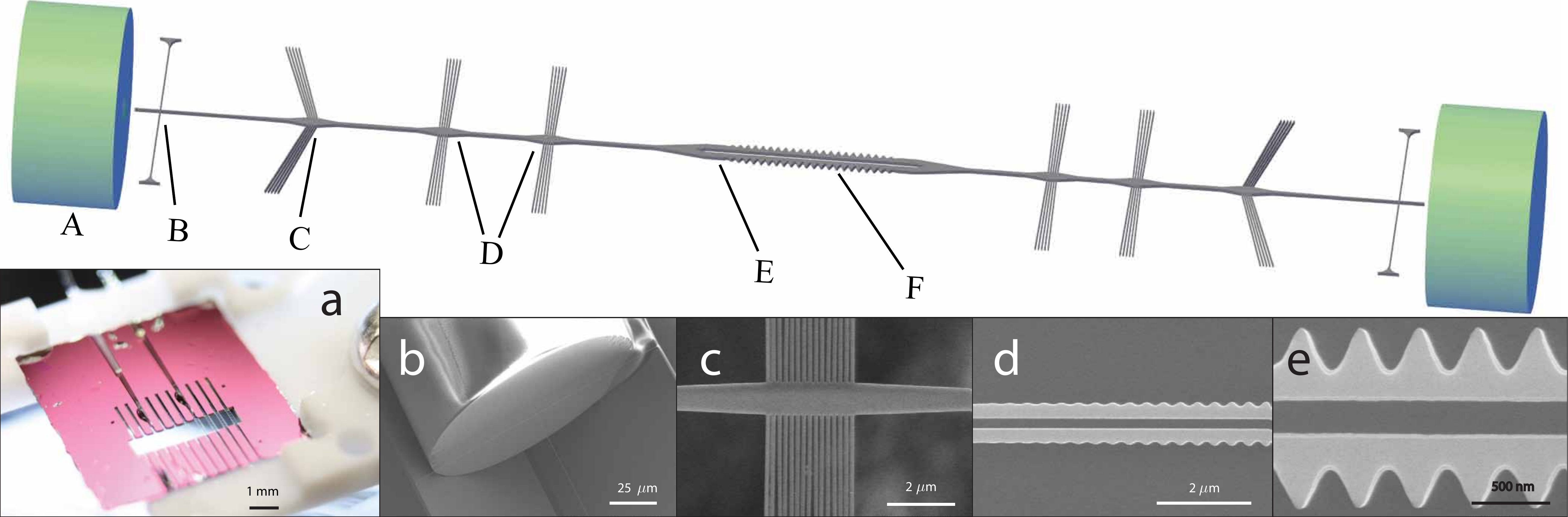}
\caption{Center - Schematic of the waveguide chip, illustrating the various regions of the waveguide.  Bottom - (a) Optical image of the fiber-coupled waveguide chip showing the through-hole for optical access.  Zoom-in SEM image of (b) the adiabatic fiber-coupling region (A), (c) the alignment, mechanical support, and thermal heat-sink tethers (B,C,D), (d)  the tapered region of the APCW (E), and (e) the central APCW region (F).  The sinusoidal modulation facilitates high-precision fabrication.  Other elements (not shown) are side thermal contacts which consist of a pair $7.5$~$\mu$m wide SiN rails extending across the entire length of the waveguide and connecting to the substrate.} \label{fig:wg_fab}
\end{center} 
\vspace{-8mm}
\end{figure*}

One strategy for trapping Cs atoms within the gap of the APCW is to use the dielectric-band mode blue-detuned from the Cs D1 line as a trapping beam, and the air-band mode as a probe on the D2 line of the trapped atoms.  In this scenario~\cite{Hung2013}, Cs atoms are trapped between the parallel dielectrics where the dielectric-band mode has an intensity null in the $x$-$y$ plane (Fig.~\ref{fig:mode_trap}a) and the Casimir-Polder (CP) force provides additional confinement in the vertical $z$-direction (Fig.~\ref{fig:mode_trap}d, e). Further vertical confinement can be provided by an additional guided mode red-detuned from the Cs D2 transition. 
For the band structure shown in Fig.~1b and with counter-propagating 30 \micro W TE-mode fields blue-detuned 30-GHz from the Cs D1 $F=4\rightarrow F'=4$ transition combined with 15 \micro W of counterpropagating TE-mode fields red-detuned 300~GHz from the D2 $F=4\rightarrow F'=5$ transition, we expect a trap depth of $\simeq 5$~mK and trap frequencies of $\{\nu_{x} = 3.5, \; \nu_{y} =1.4 , \; \nu_{z}= 0.7\}$~MHz. Here, the total power within the device is 90 \micro W.  


As shown in Fig.~3, we incorporate several elements into the waveguide structure to in- and out-couple light, to provide mechanical support, and to improve heat dissipation.  SEM images taken along the length of the SiN waveguide show the various sections of the device, including a waveguide-to-fiber coupling region (Fig.~\ref{fig:wg_fab}b), mechanical support and thermal tethers (Fig.~\ref{fig:wg_fab}c), a tapered region of the APCW (Fig.~\ref{fig:wg_fab}d), and finally the central APCW region (Fig.~\ref{fig:wg_fab}e). 

The waveguide-to-fiber coupling in Fig.~\ref{fig:wg_fab}b  consists of a slow tapering of the nanowire-waveguide from a nominal width of $300$~nm down to an endpoint near the fiber facet of width $130$~nm and provides efficient optical mode-matching to an optical fiber \cite{Cohen2013} (Nufern 780HP fiber; mode field diameter $5$~$\mu$m). To support the nanowire-waveguide, nanoscale tethers are run from the side of the waveguide either directly to the substrate or to side `rails'  of $7.5$~$\mu$m wide SiN that extend the entire waveguide length and connect to the substrate at either end of the waveguide (Fig.~\ref{fig:wg_fab}B, C, D show the tethers).  The tethers are each $90$~nm wide, and consist of a single tether for fiber alignment at the ends of the waveguide and multi-tether arrays of $15$ tethers,  spaced at a $220$~nm pitch. Finite-difference time-domain (FDTD) simulations \cite{Lumerical} show that the input coupling efficiency of the taper and single alignment tether is $\simeq 75$\% for light near the D2 line of Cs. The multi-tether supports provide anchoring against the high stresses within the device and increase device-substrate thermal contact.  Optical scattering is minimized at the multi-tether attachment points by tapering the waveguide width up to $1$~$\mu$m (see Fig.\ref{fig:wg_fab}b).  FDTD simulation shows that the scattering loss at the multi-tether points is $\lesssim 0.5\%$.

The nanowire waveguides as shown in Fig. \ref{fig:wg_fab} are formed from a thin film of stoichiometric SiN $200$~nm in thickness, grown via low-pressure chemical vapor deposition on a $\left( 100 \right)$ Si substrate of $200$~$\mu$m thickness.  This sort of SiN has exhibited low optical loss in the near-infrared~\cite{Barclay2006,Jayich2008,Wilson2009}, and large tensile stress ($\sim \!\! 1$~GPa)~\cite{TempleBoyer1998}. A $1\times 5$~mm window opened through the Si substrate provides optical access for laser trapping and cooling, with the nanowire waveguides extending across the length of window.  Even with the extreme aspect ratio of the nanowire waveguides, the high tensile stress of SiN preserves mechanical stability and alignment.

In order to obtain smooth waveguide side walls of vertical profile and to avoid damage during the SiN etch, we employ an inductively-coupled reactive-ion etch (ICP-RIE) of low DC-bias and optimized C$_4$F$_8$ and SF$_6$ gas ratios. A similar etch has been used to create record-high Q SiN micro-ring optical cavities near $800$ nm~\cite{Barclay2006,Barclay_thesis}.  Fabrication of the waveguide chip begins with a UV lithography step to define the back window region. We then use a single $e$-beam lithography step to define the fine features of the waveguide, and to set the fiber v-groove position and width (which ultimately determine the fiber-waveguide alignment). A piranha clean removes any resist residue prior to a potassium hydroxide (KOH) wet etch, which opens a through-hole in the Si substrate defined by the two SiN windows on back and front.  
After additional Nanostrip cleaning, the chip is transferred to an isopropyl alcohol solution where it is dried using a critical point drying step to prevent stiction of the double-wire APCW section.  Lastly, an O$_{2}$ plasma clean removes any residual particles on the waveguide surface.

Once fabricated, anti-reflection coated optical fibers are mounted into the input and output v-grooves in the Si substrate. The fiber-waveguide separation is set for optimal coupling (typically $\lesssim 10$~$\mu$m) before the fibers are affixed in place with UV curing epoxy.  The Si chip and fibers are then attached to a vacuum-compatible mount (see Fig.~\ref{fig:wg_fab}a) and loaded into a vacuum enclosure (reaching  $\sim 10^{-9}$ Torr) with optical fiber feedthroughs~\cite{Abraham1998}.

In order to measure the broadband reflectivity and transmission of the APCW, we utilize a broadband super-luminescent diode optical source and optical spectrum analyzer. Figure~4a shows the measured normalized reflection $R$ and transmission $T$ spectra over a frequency range of $320$-$360$~THz for a typical APCW waveguide. The measured spectra demonstrate that the fabricated APCW has the desired photonic bandgap, with the dielectric and air band-edges closely aligned with the D1 and D2 lines of Cs, respectively, and in reasonable agreement with the theoretical spectra in Fig.~4c-d.  From the average reflection level within the photonic bandgap, we estimate the total single-pass coupling from optical fiber to APCW to be $\simeq (60 \pm 5)\%$.  The high-frequency oscillatory behavior of the reflected and transmitted intensities is due to parasitic reflections from the AR-coated input fiber facet ($\sim 0.1\%$ reflection) and the input tether ($\sim 0.2\%$). Based upon previous measurements for similar waveguides, we estimate that the power loss coefficient of the unpatterned nanobeam sections is $\sim 4$~dB/cm.

\begin{figure}
\begin{center}
\includegraphics[width=\columnwidth]{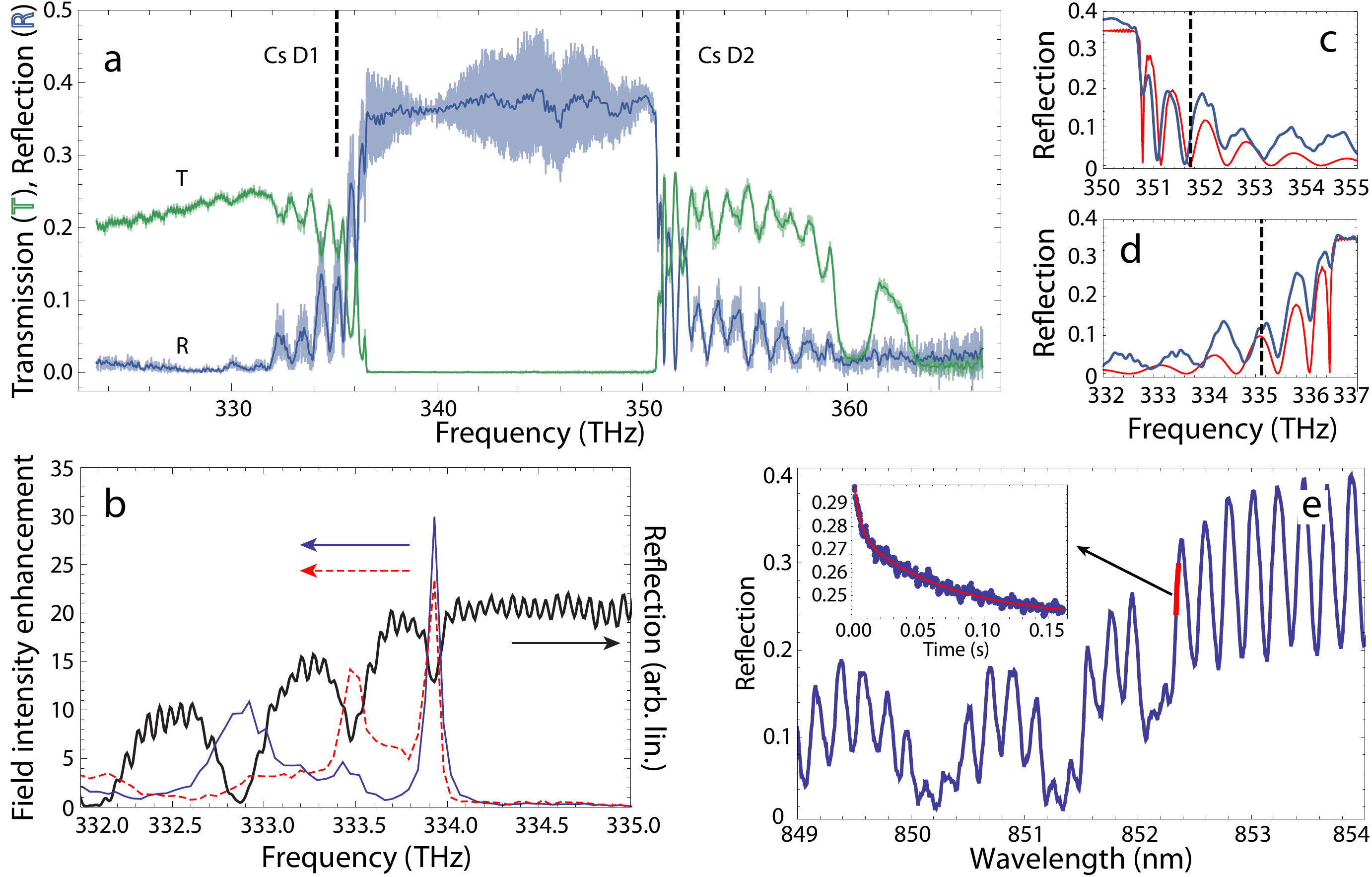}
\caption{(a) Plot of the measured reflection and transmission spectra of the complete device. A smoothing filter is applied to the raw (transparent curves) reflection measurement, yielding the solid lines. Within the bandgap, the transmitted optical power is 30 dB below the reflected power, consistent with the optical spectrum analyzer noise floor. (b) Intensity of the scattered light imaged from a region near the center (blue) and 1/3 from the end (red) of the photonic crystal section, and (black) reflected optical power as functions of the wavelength of an incident probe laser.  (c,d) Finite-difference time-domain simulation (red) and measurements (blue) of the reflection spectra near the air (c) and dielectric (d) band of the device.  (e) Thermal tuning of a waveguide reflection spectrum (red line) with respect to the device reflection spectrum (blue) and time domain response of the device reflectivity (inset) to a step function input for which the intra-device power is $\simeq 60$~\micro W. The time-domain response is fit with a double-exponential function with time constants of 70~ms and 5~ms.} \label{fig:wg_meas}
\end{center} 
\vspace{-8mm}
\end{figure}

Because of the finite length of the APCW reported in this work, the spectral regions near the bandgap exhibit slowly oscillating fringes in transmission and reflection which can be intepreted as low-finesse cavity resonances of the APCW section. FDTD simulations reproduce the oscillatory behavior, as shown in Fig.~4(c-d). The enhancement of $\Gamma_{1D}$ will be similarly oscillatory, in analogy with the Purcell effect in cavity QED (see e.g., \cite{Vahala:2003cx}). We image the scattered light directly above the waveguide as the frequency of a laser source is scanned across the slow fringes at the band-edge.  As shown in Fig.~\ref{fig:wg_meas}b (red and blue traces), the scattered radiation from the APCW section is modulated as the input frequency is scanned from resonance to antiresonance.  Based upon the measured enhancement of intensity within the APCW (normalized with respect to illumination several THz from the bandgap) of $\simeq 30$ (i.e., a cavity finesse of $\simeq 10$), we estimate that at the reflection minimum nearest the bandgap, $\Gamma_{1D}/\Gamma_{0} \simeq 20$ for a Cs atom in the $6^{2}P_{3/2}$ $|F=5, m_{F}=0\rangle$ excited state.

The optical power handling capabilities of our devices ultimately limit the trapping schemes we can consider. Figure~\ref{fig:wg_meas}e shows the time- and frequency-dependent reflection signal of a single-mode laser, with frequency tuned to the band edge of the APCW.  In this measurement, a heating laser, with frequency of 335~THz, is abruptly switched on at the input fiber port using an acousto-optic modulator at time $t=0$, and then switched off at $t=200$~ms. The increase in the temperature for a device in ultrahigh vacuum can be estimated by measuring the shift of the fast fringe shown Fig.~\ref{fig:wg_meas}a. For an input power of $95$~$\mu$W, which corresponds to an intra-device power of approximately 60 \micro W, we measure a wavelength shift of the reflection spectrum of $\delta\lambda=40$~pm.  For a thermo-optic coefficient of SiN of d$n$/d$T$$\approx 2.5\times 10^{-5}$~\cite{Barclay_thesis}, and an average energy density overlap $\eta_{E}=0.85$ of the guided mode within the SiN, this corresponds to an average temperature rise of $\delta T \simeq 2$~$^{\circ}$C.  This rise is roughly an order of magnitude smaller than the temperature rise of our waveguide devices without thermal rails.

The waveguide technology presented here represents an important step towards experiments with ultracold atoms and nanophotonic chip-based optical circuits.  The chip-based technology allows nearly full optical access for cooling and manipulation of atoms in the near-field of nanowire waveguides.  Integrated optical fibers also allow for highly efficient optical input and output channels for light coupled to arrays of atoms trapped along the slot of the APCW.  Through lithographic patterning of the nanowire waveguides, we have shown that photonic bandgaps and band-edges may be reliably produced in the vicinity of electronic transitions of atoms, a key requirement for strongly coupling atoms to the guided modes of the structure.  Indeed, current experiments~\cite{Goban:2013wp} with similar nanowire waveguides have yielded an unprecedented spontaneous emission coupling factor of $\Gamma_{1D}/\Gamma^{\prime}=0.32\pm0.08$ for one Cs atom localized near the peak of the probe mode of the APCW (Fig.~\ref{fig:mode_trap}(a)). 

Further improvements in waveguide performance resulting from lower optical absorption and scattering loss in the nanowire waveguides should enable the trapping of atoms via the fields of far off-resonant guided modes~\cite{Hung2013}.  In addition, the mechanical compliance of the suspended nanowire waveguides enables novel electro-mechanical tuning methods, similar to those recently demonstrated in tunable Si nanophotonic structures~\cite{Winger2011}.  Such fine-tuning would allow not only for precise alignment of photonic band-edges with atomic resonances, enabling strong light-matter interactions, but also dynamic atom-photon circuits in which optical dispersion of the APCW could be changed rapidly in comparison to the time scales for atomic radiative processes (e.g., photon mediated atom-atom interactions).

This work was supported by the IQIM, an NSF Physics Frontiers Center with support of the Moore Foundation, the DARPA ORCHID program, the AFOSR QuMPASS MURI, the DoD NSSEFF program (HJK), NSF PHY-1205729 (HJK), and the Kavli Nanoscience Institute (KNI) at Caltech. SPY and JAM acknowledge International Fulbright Science and Technology Awards.

\end{document}